# Electronic excitations of a single molecule contacted in a three-terminal configuration

*Edgar A. Osorio,[1] Kevin O'Neill,[1] Maarten Wegewijs,[2] Nicolai Stuhr-Hansen,[3] Jens Paaske,[3] Thomas Bjørnholm[3] and Herre S.J. van der Zant[1*]*

[1] Kavli Institute of Nanoscience, Delft University of Technology, PO Box 5046, 2600 GA, The Netherlands

[2] Institut für Theoretische Physik A, RWTH Aachen, 52056 Aachen, Germany and Institut für Festkörper-Forschung - Theorie 3, Forschungszentrum Jülich, D-52425 Jülich, Germany

[3] Nano-Science Center (Department of Chemistry and Niels Bohr Institute), University of Copenhagen, Universitetsparken 5, DK-2100, Copenhagen, Denmark.

[*]EMAIL: h.s.j.vanderzant@tudelft.nl

ABSTRACT. Low-temperature three-terminal transport measurements through a thiol end-capped $\pi$-conjugated molecule have been carried out. Electronic excitations, including zero and finite-bias Kondo-effects have been observed and studied as a function of magnetic field. Using a simplified two-orbital model we have accounted for the spin and the electronic configuration of the first four charge states of the molecule. The charge-dependent couplings to gate, source and drain electrodes suggest a scenario in which charges and spins are localized at the ends of the molecule, close to the electrodes.

Molecules offer a high chemical versatility at the nanometer length scale and a key challenge is to exploit this in single-molecule electronic devices [1-3]. Transistor effects [4-13] and memory operation [14] have been demonstrated, but the properties of single organic molecules coupled to metallic electrodes are still far from understood. In short the energetics of electron transfer to and from a molecule is determined by the intrinsic electronic spectrum of the molecule, and the electronic coupling of the molecule to the environment. The latter is defined both by classical Coulomb charging effects, which strongly depend on the polarizability of the surroundings, as well as by the direct electronic tunnel coupling to the electrodes. For small isolated molecules in the gas phase or in solution the coupling to the environment is well understood. However, the presence of two metal electrodes near the molecule renormalizes its properties and single molecules trapped between metal contacts remain poorly understood and are hence under intense investigation. So far a consistent picture has emerged for two-terminal transport through alkane chains [15] and a family of biphenyl molecules with amine linker groups [16] – both examples of electron transfer by direct tunnelling from source to drain via an off-resonance transport mechanism involving only virtual charge states of the molecule.

In three-terminal devices the gate electrode can bring molecular levels into and out of resonance with the Fermi energy of the electrodes so that different charge states (redox states) [4-13] can be accessed. Contrary to the direct off-resonance tunnelling process mentioned above, Coulomb charging effects play a significant role because the molecules accept one or more unit charges in the two step process that takes charges from the source to the molecule and subsequently from the molecule to the drain. For a weak electronic molecule-electrode coupling, this sequential tunnelling process is the most dominant transport mechanism. If the molecule is intermediately coupled to the source, and drain-electrodes, higher-order tunnelling processes also become important and one enters a new regime of correlated transport, exhibiting many-body effects such as the Kondo-effect and vibron-assisted inelastic cotunnelling. Metallic nanotube quantum dots displaying Kondo physics have become archetypal examples of such effects [17-19].

To elucidate the electronic properties of single-molecule junctions we have studied transport in electromigrated junctions [20] containing a thiol end-capped oligophenylenevinylene molecule in which five benzene rings are connected through four double bonds (OPV-5). In addition to a delocalised $\pi$-electron system this derivative has n-$C_{12}H_{25}$ side arms to make it soluble and acetyl protected thiol end-groups to ensure bonding with the gold electrodes [21] (see Fig. 1c). Electromigration was performed on a 10 nm thick gold wire with a width of 100 nm and a length of 500 nm (see also Fig. 1b) using a recently developed self-breaking method [22]. Molecule deposition was done from solution as described in Ref. [23].

Electromigration is a statistical process and in total 415 junctions with OPV-5 have been made. In Ref. [23] we have described the characteristic transport features of seven OPV-5 junctions, which all display an order of magnitude reduction of the addition energy relative to the optical (charge neutral) HOMO-LUMO gap [8], and vibrational mode spectroscopy in accordance with the Raman-vibrational fingerprint of the OPV-5 molecule. We also found that different junctions have different electronic coupling between molecule and electrodes, indicating that we can not control the exact molecule coupling. Here, we discuss in detail the measurements on one particular sample out of these seven. It exhibits intermediate coupling to source and drain electrodes so that Kondo resonances and inelastic cotunneling become visible. These features allow us to assign the spin and orbital filling to four subsequent charge states.

The transport-measurements are represented in Fig.2a as a plot of the differential conductance as a function of applied bias, and gate-voltage. Similar to the well established bias/gate-spectroscopy of semiconducting and nanotube quantum dots, we observe crossing bright lines separating regions of high conductance (single-electron tunnelling (SET) regime) from enclosed diamond-shaped black regions with very low conductance. Horizontal, slightly dimmer, lines within these regions are due to higher-order tunnelling processes.

The overall structure of the bright slanted lines reveals three crossing points, separating four different charge states of the molecule. Numbers in Fig. 2b indicate the charge states (see discussion below). Zero-bias Kondo resonances are visible in the N = +1 and N = +3 charge states; the N = +2 state reveals two finite-bias resonances located symmetrically around zero bias. The sequence: single peak, split peak, single peak is highlighted in Fig. 2b and shows a striking resemblance to data on carbon nanotube quantum dots [17,18,19]. Another characteristic feature of the Coulomb diamonds is an addition energy of 65 ± 5 meV for N = +1 and of 115 ± 5 meV for N = +2, similar to earlier reports on the same molecule [8, 23].

The crossing between N = +1 and N = +2 is of particular interest and Fig. 2c shows a high-resolution dI/dV map of this region. On the right hand-side the zero-bias Kondo peak is clearly visible; on the left-hand-side there are two peaks at a bias voltage of ±1.7 mV. Examples of differential conductance traces for both cases are plotted in Fig. 2f. The very small peaks superimposed on the main peaks are believed to be vibrational side bands at energies comparable to those measured in other samples. They are also faintly present in the SET regime.

Inspection of the data in Fig. 2c also shows an excitation running parallel to the N = +2 diamond edge which exhibits negative differential resistance (NDR as illustrated in Fig.2e, upper trace). In Fig. 2c the energy of this excitation can be read off as the distance from the zero-bias axis to the crossing point with the diamond edge; it equals 6 ± 1 meV. At the negative side this excitation is also present at the same energy as a zero-bias conductance plateau. For the N = +2 state, excitations are present at 30 ± 1 meV and ± 37 ± 1 meV.

We have studied the low-bias features of Fig. 2c as a function of temperature and magnetic field. The temperature dependence of the maximum conductance of the peaks at N=+1, +3 is consistent with the regular S = 1/2 Kondo effect as shown in Figure 3b by the red lines which are fits to the expected behaviour:

$$G_K = G_c + G_a \left[ 1 + (2^{1/s} - 1) T^2 / T_K^2 \right]^{-s} \qquad (1)$$

where s=0.22 and where $G_c$, $G_a$ are fitting parameters. The fits yield Kondo temperatures of 37 and 30 K for N=+1 and + 3 respectively, roughly consistent with the full width at half maximum of the zero-bias peaks which are 32 K for N=+1 and 24 K for N=+3. For N=+2, the zero-bias conductance first increases as temperature is raised. At 35 K it reaches a maximum and for higher temperatures the conductance decreases again. The value of 35 K corresponds to an energy of 3 meV, which is close the distance between the split low-bias peaks in this charge state (see Fig. 2f).

Figure 4a and 4d show the conductance traces at gate voltages 0.56 V (N = +1) and -0.72 V (N = +2) respectively for various magnetic fields. For N = +1 the magnetic field splits the peak in two and the magnetic field dependent data are again consistent with S = 1/2 Kondo physics. Measurements on the finite-bias peaks inside the N = + 2 diamond show that a magnetic field splits each peak into two clear peaks. At negative bias, the third derivative (Fig. 4f) shows the presence of three peaks. This evolution in a magnetic field, B, is consistent with inelastic cotunnelling from a singlet ground state to an excited triplet (cf. supporting information, section S3). The splitting in each peak should be $g\mu_B m_s B$ with $m_s = \pm 1$. With a g-factor of 2 the splitting in Fig. 4e,f yields an $m_s$-value of 0.87 at negative bias and 1.0 at positive bias, consistent with the identification of the triplet state as the excited state.

From the magnetic field measurements, we conclude that the Coulomb diamond with the split peaks in zero-magnetic field has a ground state with an even occupancy. We assign N = +2 to this state and starting from this, the electronic and spin spectrum can be identified using a model that involves two weakly interacting states with an antiferromagnetic exchange energy J. From standard Coulomb blockade theory we employ the following effective Hamiltonian for the molecule (cf. Refs [24,25]):

$$H_{mol} = E_C (\sum_{\substack{i=A,B \\ \sigma=\uparrow,\downarrow}} n_{i\sigma} - N_0 - N_G)^2 + \delta n_B + dU \sum_{\substack{i=A,B \\ \sigma'=\uparrow,\sigma=\downarrow}} n_{i,\sigma'} n_{i,\sigma} + J \vec{S}_A \cdot \vec{S}_B \qquad (2)$$

where $\{n_{A\uparrow}, n_{A\downarrow}, n_{B\uparrow}, n_{B\downarrow}\}$ denote the occupation operators for two distinct orbitals $A$ and $B$ with corresponding spin ½ operators $\vec{S}_i$. The parameters represent an orbital splitting $\delta$, an electrostatic charging energy $E_C = e^2/2C$ defined in terms of a total capacitance $C$, a gate-induced charge $N_G = C_g V_g / e$, an additional *intra*-orbital Coulomb repulsion $dU$ and an *inter*-orbital antiferromagnetic exchange-coupling $J>0$.

From the Hamiltonian in Eq. 2 the addition and excitation energies can be calculated assuming that charging the molecule is merely a matter of filling up two distinct "frozen" orbitals. Fig. 5a summarizes the results for the N=+1 and N=+2 charge states. We now compare the predictions of this model with the data and check for consistency. For *N=+2* the data reveal a singlet ground state together with inelastic cotunnelling lines at ±1.7 meV ($ES^1_{N=+2}$), which are connected to a triplet state. Thus, the exchange coupling, *J*, is anti-ferromagnetic and equals 1.7 meV. The data at *N=+1* show a strong excitation at 6 ± 1 meV, which we identify as the level mismatch $\delta$ ($ES^1_{N=+1}$). For the *N=+2* state another excitation ($ES^2_{N=+2}$) is visible at ±37 meV. At the positive side there is also an additional excitation at 30 meV connected to a transition between the second *N=+2* excited state and the first *N=+1* excited state ($ES^2_{N=+2}$ - $ES^1_{N=+1}$). The difference between these excitations at 30 and 37 meV should be the level mismatch $\delta$, which is indeed the case. The value of 37 meV for the second excitation for the doubly-charged state indicates that $dU = 42 ± 2$ meV. The experimental values for the addition-energies are $E_{add,N=+1}$ = 65 ± 5 meV and $E_{add,N=+2}$ = 115 ± 5 meV. From the first number we deduce a charging energy $E_C$ = 30 ± 3 meV, whereas the reading for $E_{add,N=+2}$ implies that $E_C$ = 38 ± 4 meV. The fact that these two numbers are close serves as a positive consistency check of our two-orbital model.

We emphasize that the analysis presented above accounts for all the electronic and magnetic excitations of the OPV-5 junction throughout four consecutive charge states. The remaining question is how the two weakly interacting and spatially separated states A and B relate to the molecule in the junction. Here we note that the presence of two nearly degenerate antiferromagnetically coupled states corroborates a picture invoked in previous papers involving image charges in the metal electrodes in which charges and spins localize at either end of the molecule as schematically illustrated in Fig. 5b [8,26]. In the remainder of this letter we highlight another feature of these states which hints at their molecular origin.

The line width defining the diamond-like structures in Fig. 2a is proportional to the electronic coupling, Γ, and as this figure shows, appears to be dependent on the charge state: The diamond edges are sharper for the N = +2 state compared to N = +1 and N = +3. For the region where electron tunnelling involves 0→+1→0 transitions, a Lorentzian fit through the dI/dV vs. V line shape yields Γ≈35 meV; for the +1→+2→+1 and +2→+3→+2 transitions, Γ is 6 and 22 meV respectively. A related observation is that the gate coupling for the three different charge states is not the same. For charge state +1, +2 and +3 the gate coupling equals 0.06, 0.1, and 0.07 respectively. Thus, the N=+2 charge distribution on the molecule is more susceptible to the gate field. Apparently, the wave functions are located more to the middle of the junction in agreement with the lower electronic coupling for this charge state.

By considering the plausible valance bond structures of OPV-5, we have been able to link the observed change with charge state in couplings to electrodes and gate to its molecular structure. For the first three positively charged states, these bond structures are depicted in Fig. 6. The singly charged state (top) is a spin doublet with a strongly localized charge at one terminal. The spin remains localized to the same terminal benzene ring to preserve aromaticity (undisturbed benzene rings which are particularly stable) in the remaining four rings. Charge transport could occur by charges tunnelling directly to and from this terminal state that hence defines the values of $\Gamma$ and the gate coupling.

The likely representation of the doubly charged state is shown in the middle of the figure where the ground state is represented by two states similar to the singly charged molecule, but with the possibility to mix the quinone state (blue structure in Fig. 6) into the ground state (in organic chemistry terms one would say that the state is represented by two valence bond structures in "resonance" with most weight to the black one). The admixing of the quinone state into the ground state results in a slight movement of the charges away from the electrodes resulting in a smaller $\Gamma$ and a stronger gate coupling. The triply charged state can be described as a sum of the singly and doubly charged systems and transport could again occur via a state that is very similar to the singly charged molecule (now at the left electrode), giving rise to values of $\Gamma$ and the gate coupling that are very similar to the singly charged state, as observed in the experiment.

We conclude with a few remarks on the generality and consistency of the results described in this letter. We have performed preliminary quantum chemistry calculations to account for the image charge effects and in particular the electrostatic energy a charge carrier on the molecule gains when approaching the metal electrodes. We find that this energy gain crucially depends on the exact contact geometry which involves the Au-S bond angle and the distance to the image charge. In the experiment, we therefore expect a spread in the measured addition energies and charging energies in agreement with the observations [8,23]. On the other hand, the exchange energy in this scenario is expected to be not very dependent on the exact contact geometry, since the distance between the two spins remains approximately the same. Measurements on the only other OPV-5 junction that displays higher-order processes are consistent with these observations. It also shows inelastic cotunnelling peaks at an energy of ±1.5 meV, close to the value reported here. This sample, however, exhibits higher addition energies and as a consequence only one degeneracy point could be resolved so that a further comparison with the model involving multiple charge states could not be performed. Further experimental and theoretical work would be helpful for a quantitative understanding of the charge-spin distribution in molecular junctions.

**Acknowledgment.** We thank Jos Seldenthuis, Yaroslav Blanter, Jos Thijssen, Walter Hofstetter, Karsten Flensberg and Per Hedegård for discussions. Financial support is obtained from the Dutch Organization for Fundamental Research on Matter (FOM), the 'Nederlandse Organisatie voor Wetenschappelijk Onderzoek' (NWO), the Danish Agency for Science, Technology and Innovation (J.P), the Danish Nanotechnology program, the Danish Research Council and from the EC FP6 program (contract no. FP6-2004-IST-003673, CANEL), NanoSci-ERA, the Helmholtz Foundation, and the Research Center Jülich (IFMIT). This publication reflects the views of the authors and not necessarily those of the EC. The Community is not liable for any use that may be made of the information contained herein.

**Supporting Information Available**: Calculation of the excitations of the effective two-orbital model (Sections S1), and the non-equilibrium transport including the NDR effect (S2). Effective model for the finite bias singlet-triplet Kondo effect (Section S3). Stability diagram after transferring the probe to a different dewar (Section S4). This material is available free of charge via the Internet at http://pubs.acs.org.

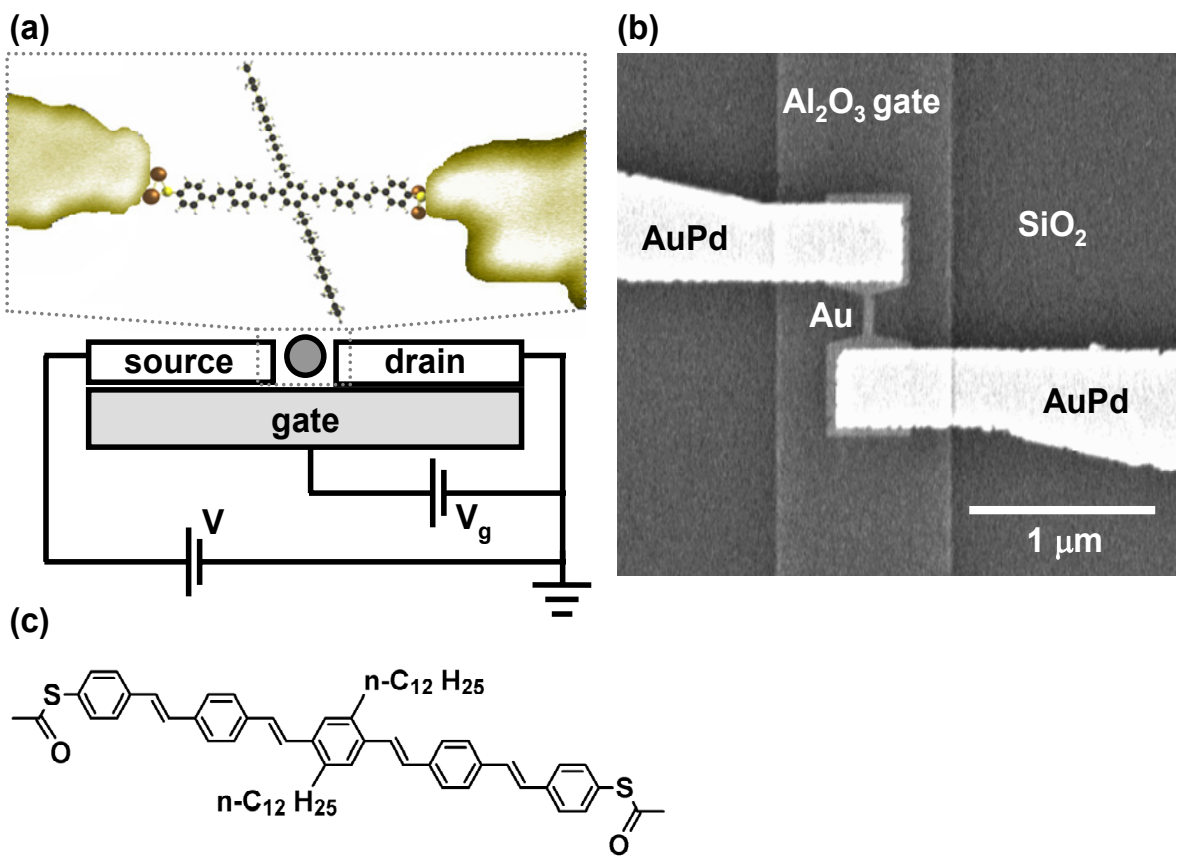

**Figure 1.** (a) Schematic device lay-out with an OPV-5 molecule bonded to two gold electrodes. (b) Fabricated device prior to breaking the small gold wire in the middle by electromigration. The junction is fabricated on top of an aluminium gate electrode, which is oxidized in air to form a 2 to 4 nm thick $Al_2O_3$ layer and at low temperatures, substantial leakage currents are typically observed for voltage above ± 4 V. Bridges are electromigrated at room temperature by ramping a voltage until a decrease in the conductance is observed, upon which the applied voltage is returned to zero [27]; the cycle is repeated until a target resistance has been reached. (c) Molecular structure of OPV-5.

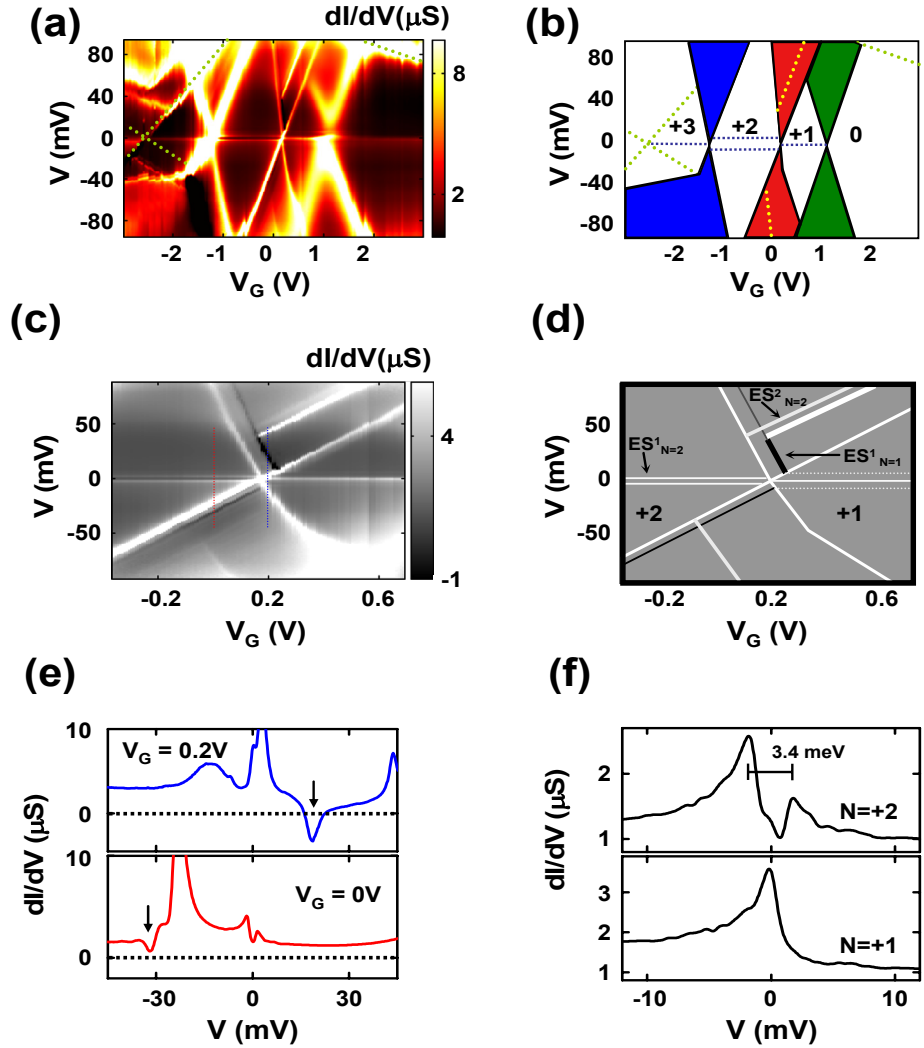

**Figure 2. (**a) Two dimensional colour plot of the differential conductance, dI/dV, versus V and $V_G$ at T = 1.7 K measured with a lock-in technique (modulation amplitude 0.4 mV). Minimum and maximum conductances are -5 µS and 25 µS respectively. Dotted green lines indicate the diamond edges of a second molecule connected in parallel. When transferring the probe into another dewar, this structure moved to a less negative gate voltage whereas the other three degeneracy points remained at the same location (see supporting information, section S4). (b) Schematic drawing of the important information contained in (a) with the charge filling sequence as discussed in the text. (c) Zoom-in between charge states +2 and +1 exhibiting respectively finite-bias and zero-bias Kondo effects. (d) Schematic drawing of (c) highlighting its important features. Dashed lines are not visible in the contrast of (c) but do appear when adjusting the contrast. It is important to note that other strong excitations do not appear,

indicating that the other electronic levels are far away or are not involved in transport. At low energy we do see lines in the second derivative, which we identify as vibrational modes. (e) Two dI/dV traces taken at different gate voltages; their position is indicated in (c) by thin vertical lines. The blue curve shows a negative differential resistance (NDR) effect at positive bias voltage. (f) dI/dV traces taken in the middle of the Coulomb diamonds of the charge states N = +2 and N = +1.

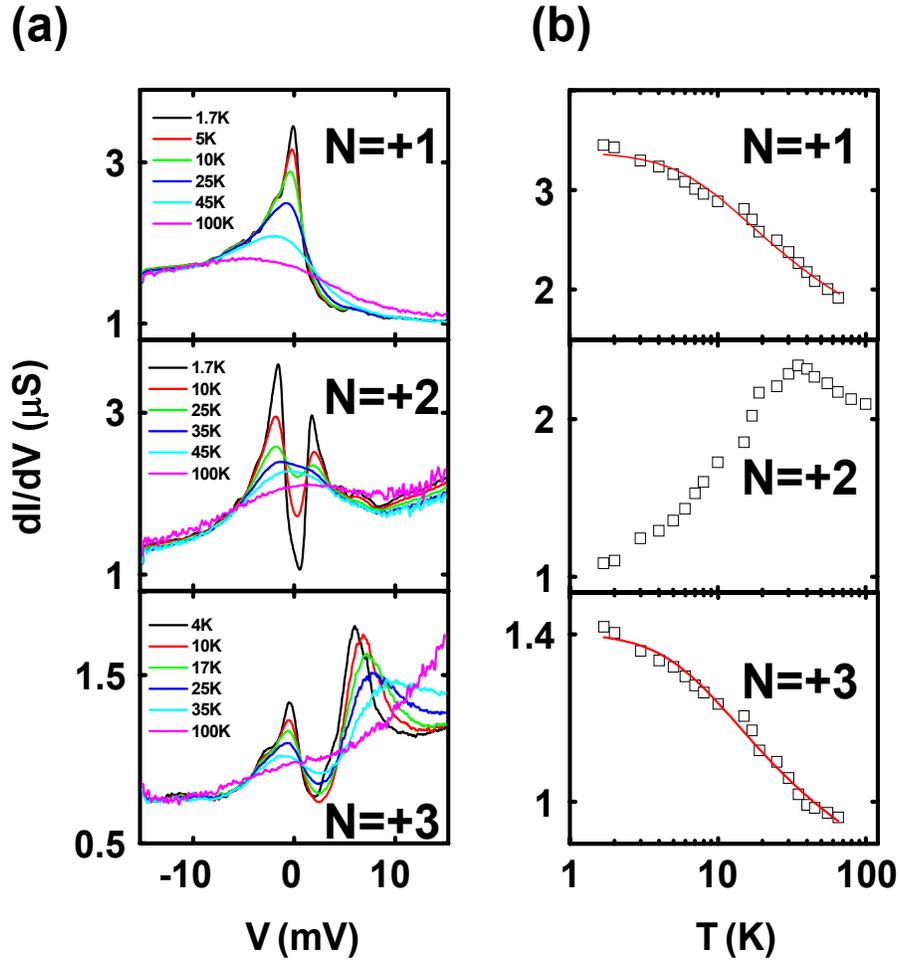

**Figure 3.** (a) dI/dV versus V at fixed gate voltage for various temperatures with $V_G$=0.56 V, -0.72 V and -2 V for charge states +1, +2 and +3 respectively. For the charge state +3 the second peak around 7 mV corresponds to the diamond edge. (b) Zero-bias conductance as a function of temperature. Lines are fits to Eq. 1 with s=0.22: the fitting parameters are $T_K$= 37 K, $G_c$= 1.0 µS, $G_a$= 2.3 µS and $T_K$= 30 K, $G_c$= 0.7 µS, $G_a$= 0.7 µS for N=+1 and N=+3 respectively. For N=+2, the zero-bias conductance first increases with increasing temperature followed by a decrease for T >35 K (~3 meV).

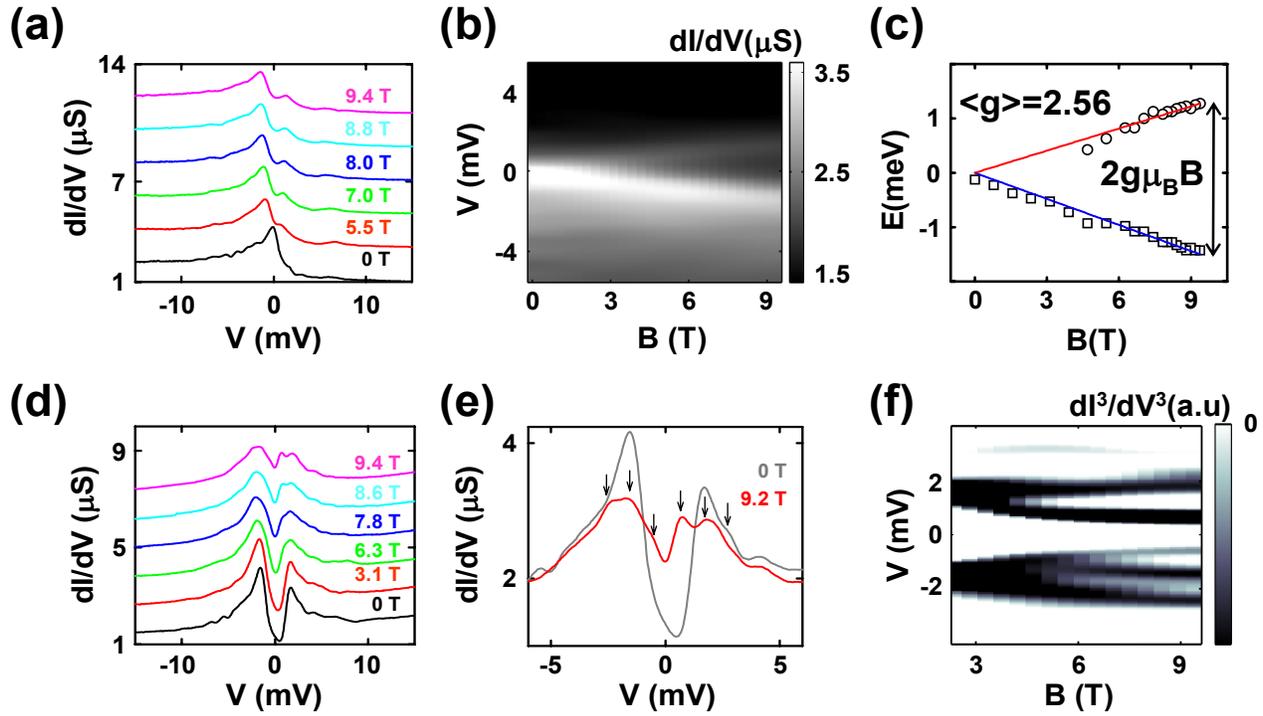

**Figure 4.** (a) dI/dV trace at $V_G = 0.56$ V (charge state N= +1) recorded for different magnetic fields as measured with a lock-in technique (modulation amplitude 0.04 mV). Traces are offset by 2 µS for clarity. (b) Grey-scale plot of dI/dV versus V and B at $V_G = +0.56$ V. (c) Peak positions as a function of magnetic field taken from the data in (b) indicating a g value of 2.56. (d) Same as (a) but at $V_G = -0.72$ V (charge state N = +2). Subsequent traces are offset by 1.2 µS. (e) dI/dV traces at $V_G = -0.72$ V measured in a magnetic field of B = 0 T (grey) and B = 9.2 T (red). Arrows indicate the predicted positions of peaks at B = 9.2 T assuming the triplet splits according to $g\mu_B B m_S$ with g = 2 and $m_S$ = -1,0,1. (f) Grey-scale plot of $dI^3/dV^3$ versus V obtained by numerical differentiation of the measured dI/dV at $V_G = -0.72$ V. The third derivative underlines the presence of three (two) peaks for negative (positive) bias voltage (peaks in the first derivative correspond to dips in the third derivative).

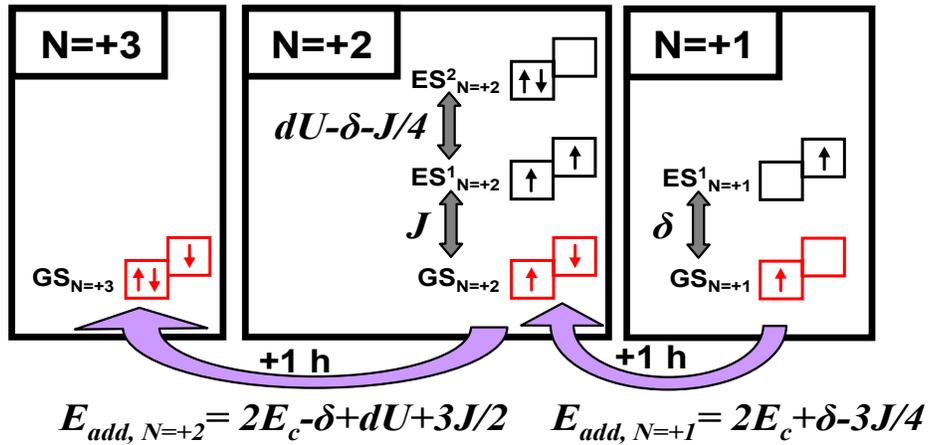

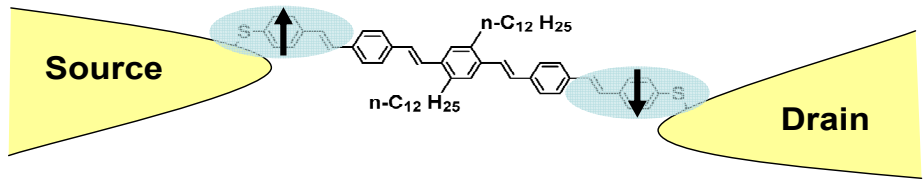

**Figure 5.** (a) Two orbitals, A and B, (depicted by the two boxes) are separated in energy by $\delta$. $GS_{N=i}$ and $ES_{N=i}$ denote ground and excited states respectively for charge state i = +1, +2 and +3 (excited states for N=+3 are not shown). J is the inter-orbital, anti-ferromagnetic exchange energy, $E_C=e^2/2C$ is the electrostatic charging energy and dU is the intra-orbital Coulomb repulsion. At the bottom, $E_{add,\ N=+i}$ denote the addition energies for charge state i=+1, +2. (b) Schematic impression of the two-site orbital model for an OPV-5 molecule connected to gold electrodes on either side.

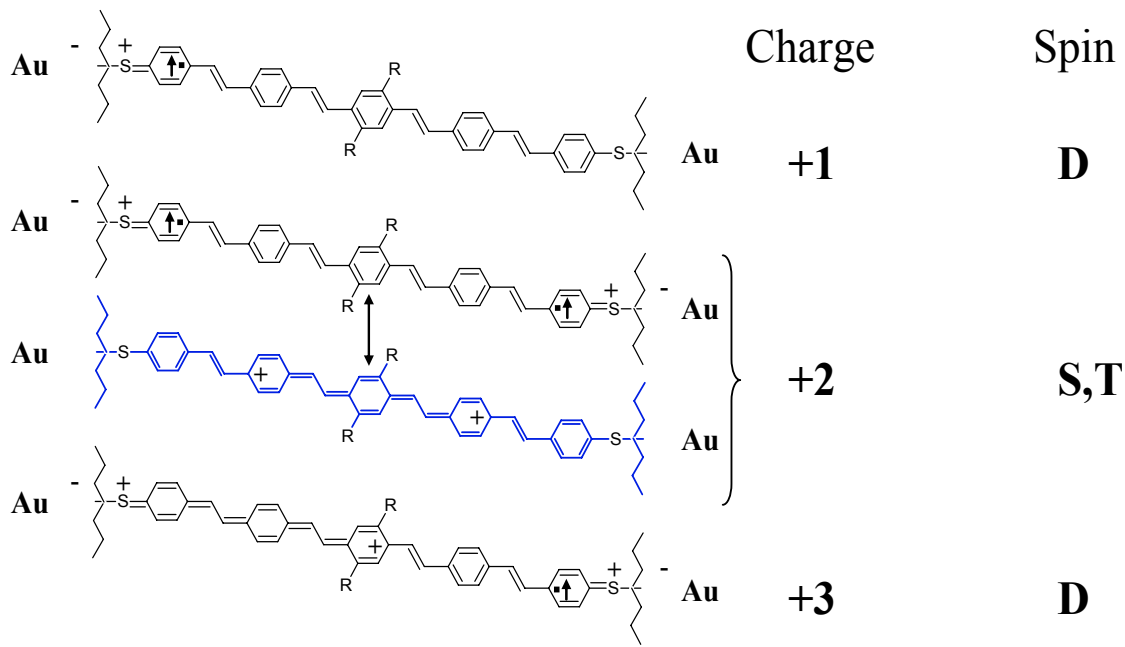

**Figure 6.** Plausible valence bond representations of the first three positively charged states. The likely spin multiplicity is indicated in the right column. The classical valence bond description of the doubly charged OPV-5 molecule involves a spinless bi-polaron structure (blue structure) in which a quinone structural motif separates the two positive charges defining a distance between the positive charges which reflects the equilibrium between the attractive structural relaxation of the molecular backbone (the quinoid structure) and the repulsion between the positive charges. The double arrow connects "resonance" structures in organic chemical terminology.

SYNOPSIS TOC (Word Style "SN_Synopsis_TOC").

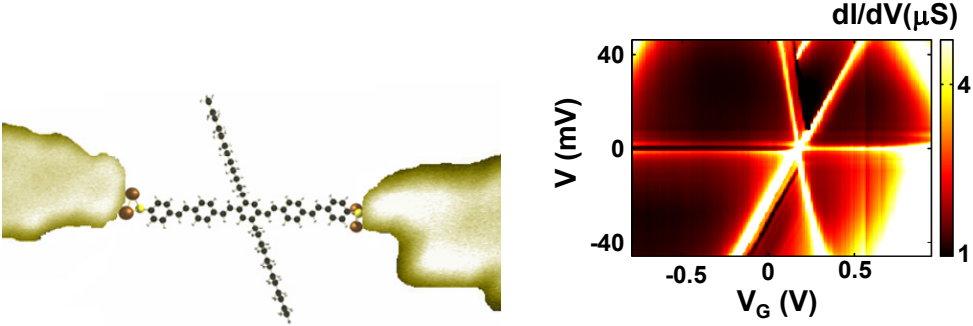